\documentstyle[12pt]{article}
\begin{document}
{\hfill PUTP-94-06}

\vspace{10mm}
\centerline{\large\bf Hadronic Matrix Elements and
Radiative ~$B\longrightarrow K^{\star}\gamma$ Decay}

\vspace{10mm}
\centerline{Jian Tang and Jing-Hua Liu}
\centerline{\small\it Department of Physics, ~Peking University,
         ~Beijing 100871, P.R.China}
\centerline{Kuang-Ta Chao}
\centerline{\small\it Center of Theoretical Physics, CCAST(World
Laboratory), Beijing 100080}
\centerline{\small\it Department of Physics, ~Peking University,
         ~Beijing 100871, P.R.China}

\vspace{10mm}
\begin{abstract}

Within the standard model, we calculate the radiative
$B\longrightarrow K^{\star}
\gamma$ decay rate based on a Bethe-Salpeter description
for the meson wave
functions and the hadronic matrix elements. With a reasonable choice
of parameters the branching
ratio BR($B\longrightarrow K^{\star}\gamma$) is found to be
$(3.8-4.6)\times 10^{-5}$, which is in agreement with the
{\it CLEO} experimental data.
We also find with $m_{b}=5.12 \ GeV$ the ratio
$R\equiv {\Gamma (B\longrightarrow K^{\star}\gamma )}/{\Gamma
(b\longrightarrow s\gamma )}=(10-12)\%$, which can be slightly
larger if a smaller $m_{b}$ is chosen.
In this approach, the light degrees of freedom in mesons are treated
as light constituent quarks with relativistic kinematics, and the form
factors in the decay amplitude are essentially determined by the
relativistic kenematics and the overlap of wave functions of the initial
and final state mesons.
Due to the large recoil momentum of the $K^{\star}$
meson at the $B$ meson rest frame, the form factors
are sensitive to the overlap
integral of the meson wave functions, which are determined
dynamically by a QCD-motivated inter-quark potential.
Relativistic effects on
the meson wave functions mainly due to
the Breit-Fermi interactions
are found to be significant in determining the decay rate.
\end{abstract}

\vspace{25mm}
\centerline{\large\bf I. INTRODUCTION}

\bigskip
The interest in studying rare $B$ decays lies with the fact that
these decays,
induced by the flavor changing $b\longrightarrow s \gamma$ neutral
currents,
are controlled by the one-loop electromagnetic penguin diagrams.
  ~They play important
roles in testing loop effects in the standard model $SU(3)_{C}\times
SU(2)_{L}
\times U(1)$ and in searching for the so-called {\it ``New Physics''}
because they involve many
important standard model parameters such as the top quark mass
$m_{t}$ and the
Cabbibo-Kobayashi-Maskawa matrix elements $\left |V_{tb} \right |$
and
$\left |V_{ts} \right |$.

It has been shown that at the quark level the inclusive decay
$b\longrightarrow s \gamma$
has a sizable enhancement due to the QCD
corrections${}^{\cite{1}}$. ~However, it is unfortunate
that we can only observe the exclusive channels, such as
$B\longrightarrow K^
{\star}\gamma$, which are hadron transitions and therefore plagued
by uncertainties in determining the
weak hadronic form factors. ~Recently the {\it CLEO} Collaboration
reported the
result${}^{\cite{2}}$
\begin{equation}
\label{a1}
BR (B\longrightarrow K^{\star}\gamma) = (4.5\pm 1.5\pm 0.9)\times
10^{-5},
\end{equation}
which makes it possible to test various models including the standard
model and
models involving new physics, provided that the estimate of hadronic
matrix elements
is under control.

By now, there have been many methods to calculate the hadronic
matrix elements,
such as the nonrelativistic quark model${}^{\cite{3}}$
${}^{\cite{4}}$ ${}^{\cite{5}}$, the HQET( Heavy Quark Effective
Theory) where the strange quark $s$ is assumed
to be heavy${}^{\cite{6}}$, the HQET combined with chiral
symmetries where the $s$ is considered
light${}^{\cite{7}}$, and the QCD sum rules${}^{\cite{8}}$. The
predicted value of
$R\equiv {\Gamma ( B\longrightarrow K^{\star}\gamma)}/{\Gamma
( b\longrightarrow s\gamma)}$ is very different among
various models, ranging from 4.5\% ${}^{\cite{5}}$ to 40\%
${}^{\cite{8}}$. Theoretically, the difficulty is mainly due
to the large recoil momentum of the $K^{\star}$ meson
in this $B$ decay process.

As an attempt to tackle this problem and, in particular,
to incorporate
the relativistic effects of the underlying dynamics, in this paper
we will use the Bethe-Salpeter description for hadronic form factors
and meson wave functions${}^{\cite{9}}$ ${}^{\cite{10}}$ to
calculate the hadronic matrix elements involved
in this rare $B$ decay.
In this approach, the light degrees of freedom in mesons are treated
as light constituent quarks with relativistic kinematics, and the form
factors in the decay amplitude are essentially determined by the
relativistic kenematics and the overlap of wave functions of the initial
and final state mesons. The form factors may then be related to the
inter-quark dynamics through the meson wave functions.
This method might allow us to calculate the form factors
in a wide range of values of the squared momentum transfer,
$q^2$, provided that a good understanding about the covariant decay
amplitude and BS wave functions is achieved.
The remainder of this paper is organized as follows. ~In Sec.II we
review the method we use. In Sec.III we apply our method to
calculate the rate of the radiative decay
$B\longrightarrow K^{\star}\gamma$ and compare it
with those given by the nonrelativistic Schr\"odinger equation and
the na\"{\i}ve scaling law
. ~Conclusions are given in the last section.

\bigskip
\bigskip
\centerline{\large\bf II. FORMALISM}

\bigskip
The standard model $SU(3)_{C}\times SU(2)_{L}\times U(1)$ has
gained great successes
and it has now been thought as
the fundamental theory to describe the weak-electromagnetic and
strong interactions.
However, there is one problem which frequently obscures theoretical
predictions
in the standard model.
That is the so-called nonperturbative(long distance) effects
that cannot
be delt with effectively at present.

Nonperturbative effects play significant roles in
the weak decays
of the heavy flavor mesons. Dealing with the hadronic matrix
elements in these
decays, one assumes that perturbative short distance($\sim\frac
1{M_{W}}$) effects and
nonperturbative long distance ($\sim\frac 1{\Lambda_{QCD}}$)
effects can be treated separately.
Perturbative effects can be calculated using the well known short
distance
techniques of QCD as modifications to the weak Halmitonian.
Nonperturbative
effects(including the exchange of soft gluons, the creation of
quark-antiquark
pairs from vacuum and the final state interactions) are absorbed
into the initial
and final hadronic states. When dealing with the hadronic matrix
elements one must have a good mastery of the hadron wave functions
connected
with the QCD nonperturbative dynamics. Unfortunately, at present
nonperturbative effects
are still difficult to be calculated from first principles.
Therefore one
often relies on various
phenomenological models such as the Bauer-Stech-
Wirbel(BSW)model${}^{\cite{11}}$, the Isgur-Scora-Grinstein-Wise
nonrelativistic quark model${}^{\cite{3}}$, etc.

In quantum field theory, a basic description for the bound states
is the Bethe-Salpeter
equation${}^{\cite{9}}$ ${}^{\cite{10}}$.
Define the Bethe-Salpeter wave function of the bound state $\mid
P\rangle $ with an overall momentum $P$ of
a quark $\psi(x_1) $ and an antiquark $\overline{\psi }( x_2) $
\begin{equation}
\label{a4}
\chi (x_1,x_2) =\langle 0\mid T\psi ( x_1)
\overline{\psi }( x_2)\mid P \rangle ,
\end{equation}
where T represents time-order product, and
transform it into the momentum space
\begin{equation}
\label{a5}
{\chi_P(q)}={e^{-iP\cdot X}\int d^{4}x e^{-iq\cdot x} \chi
(x_1,x_2)} .
\end{equation}
Here we use the standard center of mass and relative variables
\begin{equation}
\label{aa}
X= \eta_1 x_1+ \eta_2 x_2, ~~~~x=x_1-x_2,
\end{equation}
where $\eta_i = \frac {m_i}{(m_1+m_2)}(i=1,2)$.
Then in momentum space the bound state BS equation reads
\begin{equation}
\label{a6}
{(\not\!p_1-m_1)\chi_P(q)(\not\!p_2+m_2)}
= {\frac i{2\pi}}\int d^4k G(P,q-k)\chi_P(k) ,
\end{equation}
where $p_{1}$ and $p_{2}$ represent the momenta of quark and
antiquark
respectively, $G(P,q-k)$ is the interaction kernel which dominates
the inter-quark dynamics.
According to Eq.~(\ref{aa}) we have
\begin{equation}
\label{bb}
p_1= \eta_1 P+q, ~~~~p_2= \eta_2 P-q .
\end{equation}
Note that in Eq.~(\ref{a6}) $m_{1}$ and $m_{2}$ represent the effective
constituent quark masses so that
we could use the effective free propagators of quarks instead of
the full propagators.
This is an important approximation and simplification for light
quarks.
Furthermore, because of the lack of a fundamental description
for the
nonperturbative QCD dynamics, we have to make some
approximations for the
interaction kernel of quarks.

i) To solve Eq.~(\ref{a6}) one must have a good command of the
potential between
two quarks. However, the reliable information about the
potential only comes from the
lattice QCD result, which shows that the potential for a
heavy quark-antiquark
pair Q\=Q in the static limit is well described by a long-ranged
linear confining
potential ( Lorentz scalar $V_S$ ) and a short-ranged one gluon
exchange potential
( Lorentz vector $V_V$ ), i.e,${}^{\cite{15}}$
\begin{equation}
\label{a7}
{V_S(\stackrel{\rightharpoonup }{r})}=\lambda r ,
  ~~~~{V_V(\stackrel{\rightharpoonup }{r})}=-\frac 43{\alpha_s(r)
\over r} ,
\end{equation}
The lattice QCD result for the Q\=Q potential is strongly
supported by the
heavy quarkonium spectroscopy including both spin-independent and
spin-dependent effects.
In the next section we will employ the potential below regardless of
whether
the quarks are heavy or not
\begin{eqnarray}
\label{a9}
&&{V(r)}={V_S(r)+\gamma_{\mu}\otimes\gamma^{\mu} V_V(r)},\nonumber \\
&&{V_S(r)}={\lambda r\frac {(1-e^{-\alpha r})}{\alpha
r}},\nonumber \\
&&{V_V(r)}=-{\frac 43}{\frac {\alpha_{s}(r)} r}e^{-\alpha r},
\end{eqnarray}
where the introduction of the factor $e^{-\alpha r}$ is to avoid
the infrared(IR) divergence and also to incorporate
the color screening
effects of the dynamical light quark pairs on the ``quenched'' Q\=Q
potential
${}^{\cite{16}}$. It is clear that when $\alpha r\ll 1$
the potentials given in
(\ref{a9}) become identical with that given in (\ref{a7}).
In momentum space
\begin{eqnarray}
\label{a12}
&&G( \stackrel{\rightharpoonup }{p})=G_S( \stackrel{%
\rightharpoonup }{p}) +\gamma_{\mu}\otimes \gamma^{\mu}
G_V( \stackrel{\rightharpoonup
}{p}),\nonumber \\
&&G_S( \stackrel{\rightharpoonup }{p})=-\frac \lambda \alpha
\delta ^3( \stackrel{\rightharpoonup }{p})+\frac \lambda {\pi
^2}\frac 1{( \stackrel{\rightharpoonup }{p}^2+\alpha ^2)
^2},\nonumber \\
&&G_V( \stackrel{\rightharpoonup }{p})=-\frac 2{3\pi^2}
\frac {\alpha_{s}(\stackrel{\rightharpoonup
}{p})}{\stackrel{\rightharpoonup }{p}^2+\alpha ^2},
\end{eqnarray}
where $\alpha_{s}(\stackrel{\rightharpoonup }{p})$ is the well known
running
coupling constant and is assumed to become a constant of $O(1)$ as
${\stackrel{\rightharpoonup }{p}}^2\rightarrow 0$
\begin{equation}
\label{a15}
\alpha _s( \stackrel{\rightharpoonup }{p}) =\frac{12\pi }{27}%
\frac 1{\ln ( a+\frac{\stackrel{\rightharpoonup }{p}^2}{\Lambda
_{QCD}^2%
}) }.
\end{equation}
The constants $\lambda$, $\alpha$, $a$, and $\Lambda_{QCD}$ are
the parameters
that characterize the potential.
In the computation in the next section we will use
\begin{equation}
\label{a16}
{\lambda=0.183 GeV^{2}, ~~\alpha=0.06
GeV,~~a=e=2.7183,~~\Lambda_{QCD}=0.15 GeV},
\end{equation}
and will discuss the sensitivity of the results to the values of
parameters later.

ii) In solving Eq.~(\ref{a6}), in order to avoid the
notorious problem due to
the excitation of the relative time variable we have to employ the
``instantaneous
approximation''. Meanwhile, we will neglect
the negative energy projectors in the quark propagators because in
general the negative energy projector only
contributes to quantities of higher orders in
$\frac 1{m_{Q}}$, where $m_{Q}$
represents the mass of the heavy quark.

We write therefore in the ``instantaneous approximation''
\begin{equation}
\label{a17}
( \not\!p_1-m_1) \chi _P(q) ( \not\!p%
_2+m_2) =\frac i{2\pi }\int d^4k\widetilde{G}(
\stackrel{\rightharpoonup }{P},
\stackrel{\rightharpoonup }{q}-\stackrel{\rightharpoonup }{k}) \chi
_P(k),
\end{equation}
where $\widetilde{G}(\stackrel{\rightharpoonup
}{P},\stackrel{\rightharpoonup }{q}
-\stackrel{\rightharpoonup }{k})$ represents the
``instantaneous'' part of the potential $G(P,q-k)$.
This suggests the derivation of an equation for the three dimensional
BS wave function
\begin{equation}
\label{a18}
\Phi _{\stackrel{\rightharpoonup }{P}}( \stackrel{\rightharpoonup
}{q}%
) =\int dq^0\chi _P( q^0,\stackrel{\rightharpoonup }{q})
\end{equation}
by dividing both sides of Eq.~(\ref{a17}) by the propagators of two
quarks,
then integrating
over $k^0$ and neglecting the negative energy projectors
\begin{equation}
\label{a19}
\Phi _{\stackrel{\rightharpoonup }{P}}( \stackrel{\rightharpoonup
}{q}%
) =\frac 1{P^0-E_1-E_2}\Lambda _{+}^1\gamma ^0\int
d^3k\widetilde{G}%
( \stackrel{\rightharpoonup }{P},\stackrel{\rightharpoonup }{q}-
\stackrel{\rightharpoonup }{k})
\Phi _{\stackrel{\rightharpoonup }{P}}( \stackrel{\rightharpoonup
}{k}%
) \gamma ^0\Lambda _{-}^2,
\end{equation}
where
\begin{eqnarray}
\label{a20}
&&\Lambda _{+}^1=\frac 1{2E_1}( E_1+\gamma
^0\stackrel{\rightharpoonup }{%
\gamma }\cdot \stackrel{\rightharpoonup }{p_1}+m_1\gamma
^0),\nonumber \\
&&\Lambda _{-}^2=\frac 1{2E_2}( E_2-\gamma
^0\stackrel{\rightharpoonup }{%
\gamma }\cdot \stackrel{\rightharpoonup }{p_2}-m_2\gamma ^0),
\end{eqnarray}
are the remaining positive energy projectors of
the quark and antiquark
respectively.
Here, $E_1= \sqrt{m_1^2+\stackrel{\rightharpoonup }{p_1}^2},
 ~~E_2= \sqrt{m_2^2+\stackrel{\rightharpoonup }{p_2}^2}$.
{}From Eq.~(\ref{a19}) it is easy to see that
\begin{eqnarray}
\label{a22}
&&\Lambda _{+}^1\Phi _{\stackrel{\rightharpoonup }{P}}(
\stackrel{%
\rightharpoonup }{q}) =\Phi _{\stackrel{\rightharpoonup }{P}}(
\stackrel{\rightharpoonup }{q}),\nonumber \\
&&\Phi _{\stackrel{\rightharpoonup }{P}}( \stackrel{\rightharpoonup
}{q}%
) \Lambda _{-}^2=\Phi _{\stackrel{\rightharpoonup }{P}}(
\stackrel{\rightharpoonup }{q}).
\end{eqnarray}
Considering the constraint of Eqs.~(\ref{a22}),
and the requirement of space reflection of
$\gamma^0\Phi_{\stackrel{\rightharpoonup }{P}}
(\stackrel{\rightharpoonup }{q})\gamma^0 = -
\Phi_{-\stackrel{\rightharpoonup }{P}}
(-\stackrel{\rightharpoonup }{q})$
for a negative parity meson, and the constraint of the
general form of the meson
wave funstion in the rest frame (see Eq.(19) below),
then the wave function
${\Phi_{\stackrel{\rightharpoonup }{P}}}
(\stackrel{\rightharpoonup }{q})$ in a moving frame
can be written as follows${}^{\cite{9}}$
\begin{eqnarray}
\label{a24}
&&\Phi _{\stackrel{\rightharpoonup }{P}}^{0^{-}}( \stackrel{%
\rightharpoonup }{q}) =\Lambda _{+}^1\gamma ^0(
1+\frac{\not\!{P%
}}M) \gamma _5\gamma ^0\Lambda _{-}^2\varphi
_{\stackrel{\rightharpoonup }{P}}
(\stackrel{\rightharpoonup }{q})
=\frac{{\not\!p}_1+m_1}{2E_1}%
(1+\frac {\not\!P}{M})\gamma_5 \frac{{\not\!p}_2-
m_2}{2E_2}\varphi_{\stackrel{\rightharpoonup }{P}}%
(\stackrel{\rightharpoonup }{q}),\nonumber \\
&&\Phi _{\stackrel{\rightharpoonup }{P}}^{1^{-}}( \stackrel{%
\rightharpoonup }{q}) =\Lambda _{+}^1\gamma ^0(
1+\frac{\not\!{P%
}}M) \not\!e\gamma ^0\Lambda _{-}^2f_{\stackrel{\rightharpoonup
}{P}}
(\stackrel{\rightharpoonup }{q})
=\frac{{\not\!p}_1+m_1}{2E_1}%
(1+\frac {\not\!P}{M})\not\!e\frac{{\not\!p}_2-
m_2}{2E_2}f_{\stackrel{\rightharpoonup}{P}}(\stackrel{\rightharpoonup }{q}),
\end{eqnarray}
where
$\Phi _{\stackrel{\rightharpoonup }{P}}^{0^{-}}( \stackrel{%
\rightharpoonup }{q})$ and
$\Phi _{\stackrel{\rightharpoonup }{P}}^{1^{-}}( \stackrel{%
\rightharpoonup }{q})$ are the three dimensional
BS wave functions of the $0^{-}$ meson and
$1^{-}$ (S-wave) meson respectively. $P^{\mu}$ and $M$ are
the 4-momentum and mass of the meson.
$\not\!e={\gamma_{\mu}e^{\mu}}$, $e^{\mu}$ is the
polarization vector of
$1^{-}$ meson. $\varphi_{\stackrel{\rightharpoonup }{P}}
(\stackrel{\rightharpoonup }{q})$ and
$f_{\stackrel{\rightharpoonup }{P}}(\stackrel{\rightharpoonup
}{q})$ are scalar
functions of $\stackrel{\rightharpoonup }{P}$ and
$\stackrel{\rightharpoonup }{q}$ in general.
Note that in Eq.(17) the appearance of the positive energy
projectors for quark and antiquark in the meson wave functions at
the moving frame is an immediate consequence
of Eq.(16).
It is easy to see that if taking the heavy quark limit
$m_1\rightarrow \infty$, then ${p_1}^{\mu}\rightarrow
P^{\mu}$, Eq.~(\ref{a24}) becomes
\begin{eqnarray}
\label{a25}
&&\Phi _{\stackrel{\rightharpoonup }{P}}^{0^{-}}( \stackrel{%
\rightharpoonup }{q}) =\frac 1{v^0}(
1+{\not\!{v%
}}) \gamma _5\gamma ^0\Lambda _{-}^2\varphi
_{v}
(\stackrel{\rightharpoonup }{q}),\nonumber \\
&&\Phi _{\stackrel{\rightharpoonup }{P}}^{1^{-}}( \stackrel{%
\rightharpoonup }{q}) =\frac 1{v^0}(
1+{\not\!{v%
}}) \not\!e\gamma ^0\Lambda _{-}^2f_{v}
(\stackrel{\rightharpoonup }{q}),
\end{eqnarray}
where $v^{\mu}=\frac {P^{\mu}}M$, and $\varphi_v=f_v$,
which is due to vanishing color-magnetic force in the heavy quark limit.
This indicates that in the heavy quark limit the BS wave functions
respect the flavor-spin symmetry, and the light degrees of freedom
are described by $\varphi_v$, the wave function of the light
constituent quark,
which is to be determined dynamically by the BS equation.
In the rest frame of the meson
$({\stackrel{\rightharpoonup }{P}}=0) $
\begin{eqnarray}
\label{a31}
&&\Phi _{\stackrel{\rightharpoonup }{P}}^{0^{-}}( \stackrel{%
\rightharpoonup }{q}) =\Lambda _{+}^1\gamma ^0( 1+\gamma ^0
) \gamma _5\gamma ^0\Lambda _{-
}^2\varphi(\stackrel{\rightharpoonup }{q}),\nonumber \\
&&\Phi _{\stackrel{\rightharpoonup }{P}}^{1^{-}}( \stackrel{%
\rightharpoonup }{q}) =\Lambda _{+}^1\gamma ^0( 1+\gamma ^0
) \not\!e\gamma ^0\Lambda _{-}^2f(\stackrel{\rightharpoonup }{q}).
\end{eqnarray}
It is easy to show${}^{\cite{9}}$ that Eq.~(\ref{a31}) is the most
genernal form for the
$0^-$ and $1^-$ (S-wave) $q_1 {\overline {q_2}}$ meson wave
functions at the rest
frame (e.g. for the $0^-$ meson wave function there are four independent
scalar functions but with the constraint of
Eq.(16) those scalar functions can be reduced
to one and expressed exactly as Eq.(19)), and Eq.~(\ref{a24}) may be
obtained by boosting the spinors
from the rest frame to the moving frame.${}^{\cite{19}}$

Substituting Eq.~(\ref{a31}) into Eq.~(\ref{a19}),
one derives the equations for $\varphi(\stackrel{\rightharpoonup
}{q})$
and $f(\stackrel{\rightharpoonup }{q})$ in the meson rest
frame${}^{\cite{9}}$
\begin{eqnarray}
\label{a26}
M\varphi _1(\stackrel{\rightharpoonup }{q})
&=&(E_1+E_2)\varphi _1(\stackrel{\rightharpoonup }{q})\nonumber
\\
&&-\frac{E_1E_2+m_1m_2+\stackrel{%
\rightharpoonup }{q}^2}{4E_1E_2}\int
d^3k(G_S(\stackrel{\rightharpoonup }{q}
-\stackrel{\rightharpoonup }{k})-4G_V(\stackrel{\rightharpoonup
}{q}-%
\stackrel{\rightharpoonup }{k}))\varphi _1(\stackrel{\rightharpoonup
}{k})\nonumber \\
&&-\frac{(E_1m_2+E_2m_1)}{4E_1E_2}\int
d^3k(G_S(\stackrel{\rightharpoonup }{q}-%
\stackrel{\rightharpoonup }{k})+2G_V(\stackrel{\rightharpoonup
}{q}-%
\stackrel{\rightharpoonup }{k}))
\frac{m_1+m_2}{E_1+E_2}\varphi _1(\stackrel{\rightharpoonup
}{k})\nonumber \\
&&+\frac{E_1+E_2}{4E_1E_2}
\int d^3kG_S(%
\stackrel{\rightharpoonup }{q}-\stackrel{\rightharpoonup
}{k})(\stackrel{\rightharpoonup }{q}\cdot \stackrel{%
\rightharpoonup }{k})\frac{m_1+m_2}{E_1m_2+E_2m_1}
\varphi _1(\stackrel{\rightharpoonup }{k})\nonumber \\
&&+\frac{%
m_1-m_2}{4E_1E_2} \int
d^3k(G_S(\stackrel{%
\rightharpoonup }{q}-\stackrel{\rightharpoonup
}{k})+2G_V(\stackrel{%
\rightharpoonup }{q}-\stackrel{\rightharpoonup }{k}))
(\stackrel{\rightharpoonup }{q}\cdot\stackrel{%
\rightharpoonup }{k})\frac{E_1-E_2}{E_1m_2+E_2m_1}\varphi
_1(\stackrel{\rightharpoonup }{k}),
\end{eqnarray}
where
\begin{equation}
\label{a27}
\varphi _1(\stackrel{\rightharpoonup
}{q})=\frac{(m_1+m_2+E_1+E_2)(E_1m_2+E_2m_1)}{4E_1E_2(m
_1+m_2)}%
\varphi (\stackrel{\rightharpoonup }{q}),
\end{equation}
\begin{eqnarray}
\label{a28}
Mf_1(\stackrel{\rightharpoonup
}{q})&=&(E_1+E_2)f_1(\stackrel{\rightharpoonup }{q})\nonumber
\\
&&-\frac 1{4E_1E_2}\int d^3k(G_S(
\stackrel{\rightharpoonup }{q}-\stackrel{\rightharpoonup }{k})-
2G_V(%
\stackrel{\rightharpoonup }{q}-\stackrel{\rightharpoonup }{k}%
))(E_1m_2+E_2m_1)f_1(\stackrel{\rightharpoonup }{k})\nonumber \\
&&-\frac{E_1+E_2}{4E_1E_2}\int
d^3kG_S(\stackrel{\rightharpoonup }{q}-\stackrel{%
\rightharpoonup
}{k})\frac{E_1m_2+E_2m_1}{m_1+m_2}
f_1(\stackrel{\rightharpoonup }{k})\nonumber \\
&&+\frac{E_1E_2-m_1m_2+\stackrel{\rightharpoonup
}{q}^2}{4E_1E_2\stackrel{%
\rightharpoonup }{q}^2} \int
d^3k(G_S(%
\stackrel{\rightharpoonup }{q}-\stackrel{\rightharpoonup
}{k})+4G_V(%
\stackrel{\rightharpoonup }{q}-\stackrel{\rightharpoonup
}{k}))
(\stackrel{\rightharpoonup }{q}\cdot\stackrel{%
\rightharpoonup }{k})f_1(\stackrel{\rightharpoonup }{k})\nonumber \\
&&-\frac{E_1m_2-E_2m_1}{4E_1E_2\stackrel{\rightharpoonup
}{q}^2} \int d^3k(G_S(\stackrel{\rightharpoonup
}{q}-%
\stackrel{\rightharpoonup }{k})-2G_V(\stackrel{\rightharpoonup
}{q}-%
\stackrel{\rightharpoonup }{k}))
(\stackrel{\rightharpoonup }{q}\cdot\stackrel{\rightharpoonup
}{k})\frac{E_1-E_2}{%
m_2+m_1}f_1(\stackrel{\rightharpoonup }{k})\nonumber \\
&&-\frac{E_1+E_2-m_2-m_1}{2E_1E_2\stackrel{\rightharpoonup
}{q}^2} \int
d^3kG_S(\stackrel{%
\rightharpoonup }{q}-\stackrel{\rightharpoonup }{k})
(\stackrel{\rightharpoonup }{q}\cdot\stackrel{\rightharpoonup }{k})^2
\frac
1{E_1+E_2+m_1+m_2}f_1(\stackrel{\rightharpoonup
}{k})\nonumber \\
&&-\frac{m_2+m_1}{E_1E_2\stackrel{\rightharpoonup
}{q}^2} \int
d^3kG_V(\stackrel{\rightharpoonup }{q}-\stackrel{\rightharpoonup
}{k})%
(\stackrel{\rightharpoonup }{q}\cdot\stackrel{\rightharpoonup }{k})^2
\frac
1{E_1+E_2+m_1+m_2}f_1(\stackrel{\rightharpoonup }{k}),
\end{eqnarray}
where
\begin{equation}
\label{a29}
f_1(\stackrel{\rightharpoonup }{q})=-
\frac{m_1+m_2+E_1+E_2}{4E_1E_2}
f(\stackrel{\rightharpoonup }{q}).
\end{equation}

In the nonrelativistic limit for both quark and antiquark
, Eq.~(\ref{a26}) and (\ref{a28}) can be expanded in terms
of ${\stackrel{\rightharpoonup }{q}}^2/{m_1}^2$
and ${\stackrel{\rightharpoonup }{q}}^2/{m_2}^2$,
and they are identical with the Schr\"odinger equation to
the zeroth order,and with the Breit equation to the first order.
If the antiquark is light and becomes relativistic then
Eq.~(\ref{a26}) and (\ref{a28}) will include the higher order
relativistic corrections. These equations will be
solved numerically.

In the BS description the transition matrix element for $\mid
P,M\rangle \longrightarrow
\mid P^{\prime},M^{\prime }\rangle $ is given by
\begin{equation}
\label{abb}
\langle P^{\prime },M^{\prime }\mid \overline{Q^{^{\prime
}}}\Gamma Q\mid P,M\rangle
=( 2\pi ) ^4 i \int d^4p_2Tr\left\{ \overline\chi _{{P}^{\prime }}(
q^{\prime
}) \Gamma \chi _{P}(q) (\not\! p_2+m_2)\right\},
\end{equation}
where $\overline\chi _{{P}^{\prime }}( q^{\prime})=-\gamma^0
\chi ^{\dagger}_{{P}^{\prime }}( q^{\prime}) \gamma^0$.
Eq.~(\ref{abb}) can be reduced into a simpler form
expressed in terms of the
three dimensional BS wave functions,
on condition that the negative
energy projectors
in quark propagators are neglected and the kernel is
independent of the relative
time variable
\begin{equation}
\label{a33}
\langle P^{\prime },M^{\prime }\mid \overline{Q^{^{\prime
}}}\Gamma
Q\mid P,M\rangle
=( 2\pi ) ^3\int d^3p_2Tr\left\{ \Phi _{\stackrel{\rightharpoonup
}{P}^{\prime }}^{\dagger}( \stackrel{\rightharpoonup }{q}^{\prime
}) \gamma ^0\Gamma \Phi _{\stackrel{\rightharpoonup }{P}}(
\stackrel{\rightharpoonup }{q}) \right\},
\end{equation}
where the quark may change its flavor while the antiquark remains a
spectator
(see Fig. 1).
Then the normalization of the wave function $\Phi
_{\stackrel{\rightharpoonup }{P}}
( \stackrel{\rightharpoonup }{q})$ reads
\begin{equation}
\label{a30}
( 2\pi ) ^3\int d^3qTr\left\{ \Phi _{\stackrel{\rightharpoonup }{%
P}}^{\dagger}( \stackrel{\rightharpoonup }{q}) \Phi _{\stackrel{%
\rightharpoonup }{P}}( \stackrel{\rightharpoonup }{q}) \right\}
=2E=2\sqrt{M^2+\stackrel{\rightharpoonup }{P}^2}
\end{equation}

Based on the formalism above with the solutions for the BS wave
functions
, we can calculate the form factors in various
processes. We can also calculate the decay constants and the mass
differences between $0^{-}$ and $1^{-}$
mesons, etc. Along these lines a rather extensive investigation has
been made and
the results are found in agreement with the
experiments${}^{\cite{9}}$. In the next section we'll
apply this formalism to compute the decay rate of the radiative
$B\longrightarrow
K^{\star}\gamma$ decay.

\bigskip
\bigskip
\centerline{\large\bf III. The Decay $B\longrightarrow
K^{\star}\gamma$ Within the Standard Model}

\bigskip
Within the standard model, the inclusive $b\longrightarrow s\gamma$
decay
is governed
by the electromagnetic penguin operator, for $m_{s}\ll m_{b}$
${}^{\cite{12}}$
\begin{equation}
\label{a34}
{\cal H}_{eff}=Cm_b\epsilon ^\mu \overline{s}\sigma _{\mu \nu
}q^\nu (1+\gamma_5) b,
\end{equation}
where $\epsilon^{\mu}$ and $q^{\mu}$ are the polarization vector
and momentum
of the photon.
The constant C includes the QCD corrections${}^{\cite{1}}$ and the
dependence upon the
CKM matrix elements and the heavy quark masses
\begin{equation}
\label{a35}
C=\frac{G_F}{2 \sqrt{2}}\frac e{8\pi
^2}V_{ts}^{\star}V_{tb}F_2(\frac
{m_t^2}{m_W^2}),
\end{equation}
where
\begin{equation}
\label{a36}
F_2( x) =r^{-\frac{16}{23}}\left\{ \widetilde{F_2}(
x) +\frac{116}{27}\left[ \frac 15( r^{\frac{10}{23}}-1)
+\frac 1{14}( r^{\frac{28}{23}}-1) \right] \right\}
\end{equation}
with $r=\frac{\alpha _s( m_b) }{\alpha _s( m_W) }$
and $\widetilde{F_2}( x)$ given by
\begin{equation}
\label{a37}
\widetilde{F_2}( x) =\frac x{( x-1) ^3}\left[ \frac
23x^2+\frac 5{12}x-\frac 7{12}-\frac{3x^2-2x}{2( x-1) }\ln
x\right].
\end{equation}
The function $F_2( x)$ depends weakly upon the top quark mass
with a value increasing from 0.55 to 0.68 in the range
$90 GeV\leq m_t \leq 210 GeV$.

The amplitude for
$B( P) \longrightarrow K^{\star}(k,\eta )\gamma (q,\epsilon )$
is
\begin{equation}
\label{a38}
{\cal A}(B \longrightarrow K^{\star}\gamma
)=Cm_b\langle K^{\star}(k,\eta )\mid \overline{s}\sigma _{\mu \nu
}(1+\gamma_5) b\mid
B( P) \rangle \epsilon^\mu q^\nu,
\end{equation}
where $\eta$ and k are the polarization vector and momentum of the
$K^{\star}$ meson
and P is the momentum of the $B$ meson.
The hadronic matrix element involved
in this process can be expressed in term of its Lorentz structures as
follows(or equivalently with form factors
$f_1$, $f_2$ and $f_3$)${}^{\cite{12}}$
\begin{equation}
\label{a39}
\langle K^{\star}(k,\eta )\mid \overline{s}\sigma ^{\mu \nu }b\mid B(
P) \rangle =\epsilon ^{\mu \nu \alpha \beta }(A\eta _\alpha
^{\star}P_\beta +B\eta _\alpha ^{\star}k_\beta +C\eta ^{\star}\cdot
PP_\alpha k_\beta ),
\end{equation}
where the form factors A, B and C are functions of
$q^2=(P-k)^2$.
Using the Dirac matrix identity
\begin{equation}
\label{a40}
\sigma ^{\mu \nu }\gamma _5=-\frac 12i\epsilon ^{\mu \nu \alpha
\beta
}\sigma _{\alpha \beta },
\end{equation}
the matrix element of the current
$\overline{s}\sigma _{\mu \nu }\gamma _5 b$ is given by the same
form factors
\begin{equation}
\label{a41}
\langle K^{\star}(k,\eta )\mid \overline{s}\sigma ^{\mu \nu }\gamma
_5b\mid
B( P) \rangle =i[~A(\eta ^{{\star}\mu} P^\nu
-\eta ^{{\star}\nu} P^\mu
)+B(\eta ^{{\star}\mu} k^\nu -\eta ^{{\star}\nu} k^\mu )+C\eta
^{\star}\cdot P(P^\mu
k^\nu -P^\nu k^\mu )~].
\end{equation}
For the real photon these form factors are evaluated at $q^2=0$ and
the form factor C
does not contribute to this transition. The decay width
calculation is
straightforward,
\begin{equation}
\label{a42}
\Gamma (B\longrightarrow K^{\star}\gamma )=\frac{\left| C\right|
^2(m_B^2-m_{K^{\star}}^2)^3 m_b^2} {4\pi m_B^3}\left| A(0)+B(0)\right|
^2
\end{equation}

We will calculate the form factors A and B at $q^2=0$ in our BS
formalism.
It is convenient to do the computation in the rest frame of B meson
$(\stackrel{\rightharpoonup }{P}=0) $
in which the wave functions of $B^{-}(b\bar{u})$ and
${K^{\star}}^{-}(s\bar{u})$
can be expressed as follows
\begin{eqnarray}
\label{a43}
&&\Phi _B(\stackrel{\rightharpoonup
}{q})=\frac{{\not\!p}_b+m_b}{2E_b}(1+\gamma
^0)\gamma _5\frac{{\not\!p}_u-m_u}{2E_u}\varphi
(\stackrel{\rightharpoonup }{q}),\nonumber \\
&&\Phi _{K^{\star}}(\stackrel{\rightharpoonup }{q}^{\prime
})=\frac{{\not\!p}_s+m_s}{2E_s}%
(1+\frac {\not\!k}{M_{K^{\star}}})\not\!{\eta} \frac{{\not\!p}_u-
m_u}{2E_u}f(\stackrel{\rightharpoonup }{%
k},\stackrel{\rightharpoonup }{q}^{\prime }),
\end{eqnarray}
where $\stackrel{\rightharpoonup }{q}$ and
${\stackrel{\rightharpoonup }{q}}^{\prime}$
are the internal relative momentum
of B meson and $K^{\star}$ meson respectively:
$\stackrel{\rightharpoonup }{q}
=-\stackrel{\rightharpoonup }{p_u}$,
${\stackrel{\rightharpoonup }{q}}^{\prime}=
\frac{m_u}{m_s+m_u}\stackrel{%
\rightharpoonup }{k}-\stackrel{\rightharpoonup }{p_u}$.
Substituting Eqs.~(\ref{a43}) into Eq.~(\ref{a33}) we get
\begin{eqnarray}
\label{a45}
\langle K^{\star}(k,\eta )\mid \overline{s}\sigma _{\mu \nu }\gamma
_5b\mid
B(P)\rangle &=&(2\pi )^3\int d^3p_uTr\{ {\not\!{\eta}} (1+\frac
{\not\!k}{M_{K^{\star}}})
\frac{{\not\!p}_s+m_s}{2E_s}
\gamma ^0\sigma _{\mu \nu }\gamma _5\nonumber\\
&&\frac{{\not\!p}_b+m_b}{2E_b}%
(1+\gamma ^0)\gamma _5
\frac{{\not\!p}_u-m_u}{2E_u}\}
\varphi (\stackrel{\rightharpoonup
}{p_u})f(\frac{m_u}{m_s+m_u}\stackrel{%
\rightharpoonup }{k}-\stackrel{\rightharpoonup }{p_u} ),
\end{eqnarray}
where the $\bar{u}$ is a spectator in the transition, and
\begin{eqnarray}
\label{a46}
&&E_s=\sqrt{m_s^2+\stackrel{\rightharpoonup
}{p_s}^2}=\sqrt{m_s^2+(\stackrel{\rightharpoonup }{k}
-\stackrel{\rightharpoonup
}{p_u})^2},\nonumber \\
&&E_b=\sqrt{m_b^2+\stackrel{\rightharpoonup
}{p_b}^2}=\sqrt{m_b^2+\stackrel{%
\rightharpoonup }{p_u}^2},
 ~~~~~E_u=\sqrt{m_u^2+\stackrel{\rightharpoonup }{p_u}^2}.
\end{eqnarray}
Here we have employed the identity
\begin{equation}
\label{a49}
\frac{{\not\!p}_u-m_u}{2E_u}\gamma ^0\frac{{\not\!p}_u-
m_u}{2E_u}=\frac{{\not\!p}_u-m_u}{2E_u}.
\end{equation}
Considering the form factors are
independent of the polarization of
$K^{\star}$
and Lorentz indices,
we may choose some specific polarization(e.g. a
transverse one)
and Lorentz indices to do calculation.
{}From Eq.~(\ref{a45}) We obtain
\begin{eqnarray}
\label{a52}
B=-\frac {(2\pi)^3}{\eta ^ik^j-\eta ^jk^i}
\int d^3p_u&Tr&\{{\not\!{\eta}}
(1+\frac {\not\!k}{M_{K^{\star}}})
\frac{{\not\!p}_s+m_s}{2E_s}\gamma ^0\gamma ^i\gamma
^j\gamma _5
\frac{{\not\!p}_b+m_b}{2E_b}\nonumber\\
&&(1+\gamma ^0)\gamma _5\frac{{\not\!p}_u-
m_u}{2E_u}\}\varphi (%
\stackrel{\rightharpoonup
}{p_u})f(\frac{m_u}{m_s+m_u}\stackrel{%
\rightharpoonup }{k}-\stackrel{\rightharpoonup }{p_u}),
\end{eqnarray}
\begin{eqnarray}
\label{a53}
A=\frac 1{M_B}[ ~~(2\pi )^3\int d^3p_u&Tr&\{ {\not\!{\eta}}
(1+\frac {\not\!k}{M_{K^{\star}}})\frac{%
{\not\!p}_s+m_s}{2E_s}\stackrel{\rightharpoonup }{\gamma}\cdot
\stackrel{\rightharpoonup }{\eta}
\gamma _5
\frac{{\not\!p}_b+m_b}{2E_b}%
(1+\gamma ^0)\gamma _5\nonumber\\
&&\frac{{\not\!p}_u-m_u}{2E_u} \}
\varphi (\stackrel{\rightharpoonup
}{p_u})f(\frac{m_u}{m_s+m_u}\stackrel{%
\rightharpoonup }{k}-\stackrel{\rightharpoonup }{p_u})-
B\sqrt{M_{K^{\star}}^2+k^2} ~~].
\end{eqnarray}

Apart from two approximations i.e. neglecting the dependence
of the kernel on the relative time and neglecting the contribution
of higher order negative energy projectors, the expressions
(\ref{a52}) and (\ref{a53}) are rather genernal and model
independent. To calculate the values of A and B we need to know
the scalar wave functions $\varphi$ and $f$, which depend on
the dynamical model to be used. Here we will use the potential
model described in (\ref{a7})---(\ref{a16}) to solve for $\varphi$
and $f$.
Substituting $\varphi$ and $f$ obtained by solving
Eqs.~(\ref{a26}),(\ref{a27})
and Eqs.~(\ref{a28}),(\ref{a29}) into Eq.~(\ref{a52})
and Eq.~(\ref{a53}) we get
\begin{equation}
\label{a54}
\left | A(0)+B(0)\right |=0.62
\end{equation}
and
\begin{equation}
\label{a55}
BR(B\longrightarrow K^{\star}\gamma)=3.8\times 10^{-5},
\end{equation}
where we have used $\left |V_{ts} \right |=0.042$,$~m_t=150 GeV$,
$~\left |V_{tb}\right |\simeq 1$,
$~m_b=5.12 GeV$, $~m_s=0.55 GeV$, $~m_u=0.33 GeV $ and
$\tau_B\simeq 1.3ps$. Our result
is close to those of Ref.\cite{7} ($\left | A(0)+B(0)\right |=0.53$)
and Ref.\cite{13}
($\left | A(0)+B(0)\right |=0.46 $ in our notation).
Taking into account the experimental uncertainties of
$\left |V_{ts}\right |$(ranging from
0.030 to 0.054${}^{\cite{17}}$), we find the branching
ratio $BR(B\longrightarrow K^{\star}\gamma)$ ranges from
$2\times 10^{-5}$ to
$6\times 10^{-5}$ which agrees with the {\it CLEO} data
given in (\ref{a1}) within errors.
The ratio of the exclusive rate to inclusive rate, R, is expressed
in term of A and B as follows
\begin{equation}
\label{a56}
R\equiv \frac{\Gamma (B\longrightarrow K^{\star}\gamma
)}{\Gamma
(b\longrightarrow s\gamma )}\simeq \frac{m_b^3(m_B^2-
m_{K^{\star}}^2)^3}{%
m_B^3(m_b^2-m_s^2)^3}\frac 14\left| A(0)+B(0)\right| ^2
\end{equation}
Substituting Eq.~(\ref{a54}) into Eq.~(\ref{a56}) we get
\begin{equation}
\label{a57}
R=10\%,
\end{equation}

In comparision we have solved the
nonrelativistic Schr\"odinger equations for the scalar
wave functions $\varphi$ and $f$ in the nonrelativistic limit
using the same
potential as that in the BS equations
i.e. Eqs.~(\ref{a26}) --- (\ref{a29})
which are automatically reduced to
the nonrelativistic Schr\"odinger equations to the
lowest order in
${\stackrel{\rightharpoonup }{q}^2}/{m_i^2} (i=1,2)$.
We find the value of
$\left | A(0)+B(0)\right |$ is about 0.32,
significantly smaller than
that given by Eq.~(\ref{a54}).

Meanwhile, we have also assumed the scalar wave functions are
the Gaussian as in some
nonrelativistic quark model
${}^{\cite{3}}$${}^{\cite{4}}$${}^{\cite{5}}$
\begin{equation}
\label{a58}
\varphi (\stackrel{\rightharpoonup }{q})\propto e^{-{\stackrel{%
\rightharpoonup }{q}^2}/{a^2}},
 ~~~~f(\stackrel{\rightharpoonup }{q})\propto e^{-{\stackrel{%
\rightharpoonup }{q}^2}/{b^2}}.
\end{equation}
There the parameters $a$ and $b$ were obtained by the
variational method in the nonrelativistic
Schr\"odinger equation ${}^{\cite{3}}$. Here
we will use a simpler method to estimate their values.
The parameters $a$ and $b$ are connected with the mean value of
the internal momentum squared of the $B$
and $K^{\star}$ mesons
\begin{equation}
\label{a60}
a^2=\frac 43\langle \stackrel{\rightharpoonup }{q}^2\rangle _B,
 ~~~~b^2=\frac 43\langle \stackrel{\rightharpoonup }{q}^2\rangle
_{K^{\star}}.
\end{equation}
Using the ``virial theorem''
\begin{equation}
\label{a62}
\left\langle T\right\rangle =\frac 12\left\langle r
\frac{\partial V}{%
\partial r}\right\rangle
\end{equation}
and extrapolating the na\"{\i}ve scaling law obtained
from the study of
the heavy quarkonium:
$V(r)\sim C\ln r$ where $C=0.73 GeV$ ${}^{\cite{18}}$ to
the B and $K^{\star}$ mesons, we get
\begin{equation}
\label{a63}
\langle \stackrel{\rightharpoonup }{q}^2\rangle =\mu C,
\end{equation}
where $\mu $ represents the reduced mass of the meson system.
Therefore,
for $B$ meson and $K^{\star}$ meson
\begin{equation}
\label{a64}
a^2=0.30 GeV^2, ~~~~b^2=0.20 GeV^2,
\end{equation}
which are consistent with those obtained in the nonrelativistic quark
models${}^{\cite{3}}$ ${}^{\cite{4}}$ ${}^{\cite{5}}$.
Using Eqs.~(\ref{a58}), (\ref{a64}) and Eqs.~(\ref{a52}),(\ref{a53}),
we find the value of
$\left | A(0)+B(0)\right |$ is about 0.31, almost in
coincidence with our
result based on the nonrelativistivic
Schr\"odinger equation solutions.

{}From the results we obtained above we see that the decay rates given
by the relativistic
BS wave functions distinguish them significantly from those given by
the nonrelativistic Schr\"odinger
wave functions. This reflects the dynamical differences between these
two descriptions
and suggests that the relativistic effects must be considered when we
deal with the
systems containing light quarks. The scalar functions of different
desciptions are compared in Fig. 2 and Fig. 3 .
{}From Fig. 2 and Fig. 3 we see that the wave function of $B$ meson in
the BS
decription is ``fatter'' than
that in the nonrelativistic Schr\"odinger decription
and that the $0^-$ meson wave function has a longer ``tail'' than the
$1^{-}$
meson. These are relativistic effects which are mainly due
to the well known
Breit-Fermi interactions, including both spin-dependent
and spin-independent terms. In fact, in the
$0^-$ channel there is a very
short-ranged ($\delta$ function like) spin-spin force
between quarks, which is attractive and
lowers the energy level of the $0^-$ state and
pulls the quarks towards the
origin. Consequently, in momentum space the $0^-$ meson wave
function becomes
``fatter'' and has a longer ``tail''.
Whereas for the $1^-$ mesons the spin-spin force is
repulsive and weak (three times weaker than that for the
corresponding $0^-$ mesons), and is compensated
and even overwhelmed by other attractive spin-independent
relativistic corrections. These dynamical ingredients
are contained in Eq.~(\ref{a26}) for the $0^-$ meson
with Eq.~(\ref{a28}) for the $1^-$ meson, and can be
explicitly seen by the nonrelativistic reduction of these
equations in terms of ${\stackrel{\rightharpoonup }{q}}^2/{m_1}^2$
and ${\stackrel{\rightharpoonup }{q}}^2/{m_2}^2$.

In line with the observation made in Ref\cite{5}, we find that our
results depend rather strongly upon the wave functions.
When we change the ``characteristic momentum'' parameters $a$
and $b$
in Eq.~(\ref{a58}),
we find that the value of $\left | A(0)+B(0)\right |$
changes rapidly. This is
because the momentum transferred to $K^{\star}$ is very large and
$K^{\star}$ is
so far away from the zero recoil limit that the decay rate
essentially depends upon the overlap
of the wave functions of the initial and final meson states,
and in this kinematic region the overlap becomes particularly
sensitive to the broadness of the wave functions in the
momentum space. The more broad the wave functions are, the larger
the overlap and hence the decay rate.

The numerical values of branching ratio (43) and ratio $R$ (45) are
obtained by using the potential parameters (11) and quark mass
parameters given after (43). It is significant to examine the sensitivity
of the results to those parameters. In doing so, we first use the same
quark parameters as before but change the potential parameters. We find
that the value of $\left | A(0)+B(0) \right |$ is insensitive to $\alpha$
and $a$ (see (11)) (note that the screening effect of $\alpha$ is mainly
on higher excited states rather than the ground state mesons, and that
the strength of the running Coulomb
force at large distances, which is associated with $a$,
also has little effects on the ground state heavy mesons).
On the other hand, $\left |A(0)+B(0)\right |$
is increased as the string tension $\lambda$ and $\Lambda_{QCD}$
increase. This is because, larger $\lambda$ and $\Lambda_{QCD}$ result
in stronger attractive inter-quark forces, and therefore make the meson
wave funstions more compact in coordinate space and more broad in
momentum space, and hence increase the overlap integral of meson
wave functions and the value of $\left |A(0)+B(0)\right |$.
With a popular choice of $\lambda=(0.18-0.20) \ GeV^2$ and
$\Lambda_{QCD}=(0.15-0.20) \ GeV$ and other parameters unchanged,
we find instead of (43) and (45)
\begin{equation}
BR(B\longrightarrow K^{\star}\gamma)=(3.8-4.6)\times 10^{-5},
\end{equation}
and
\begin{equation}
R\equiv \frac{\Gamma (B\longrightarrow K^{\star}\gamma
)}{\Gamma
(b\longrightarrow s\gamma )}=(10-12)\%.
\end{equation}
Next, we use the same potential parameters as (11), but change the
quark mass parameters. We find that $\left |A(0)+B(0)\right |$
is insensitive to the $b$ quark mass. This is because, as naively
shown in (49), the internal momenta of quarks in a meson are mainly
determined by the reduced mass. With $m_u=0.33 \ GeV,~m_b=(4.70-5.12)
  ~GeV$, the reduced mass $\mu=(0.308-0.310)~GeV$ is almost unchanged,
and hence $\left |A(0)+B(0)\right |$ remains rather stable.
However, from (44) it is clear that the ratio $R$ is sensitive to
$m_b$. With $m_b$ going down to $4.70~GeV$ from $5.12~GeV$, $R$ will
be increased by about $30\%$ entirely due to a smaller $m_b$ in the
mass ratios in $R$,
while $R$ is decreased by less than $10\%$ due to
a slightly smaller value of $\left |A(0)+B(0)\right |$.
It is obvious that a larger value of $R$ will favour a smaller value
of $m_b$.

Very recently the $CLEO$ Collaboration has reported the first result
of inclusive $b\rightarrow s\gamma$ decay branching ratio${}^{\cite {20}}$:
\begin{equation}
BR(b\longrightarrow s\gamma)=(2.32\pm 0.51\pm 0.29\pm 0.22)\times 10^{-4},
\end{equation}
\begin{equation}
R\equiv \frac{\Gamma (B\longrightarrow K^{\star}\gamma
)}{\Gamma
(b\longrightarrow s\gamma )}=0.19\pm 0.08.
\end{equation}
Our predicted value (52) for $R$ is smaller than the experimental
value (54). However, as discussed above, if we use a smaller value
for $m_b$, the calculated value for $R$ can be increased by, say,
about $(10-20)\%$. Nevertheless, it seems to be difficult for our model to
reach the large central value of $R=0.19$ measured by $CLEO$.
Further investigations are still needed.

\bigskip
\bigskip
\centerline{\large\bf IV. SUMMARY and CONCLUSION}

\bigskip
In this paper we solve the BS equations for mesons and employ the
covariant form for the wave
functions and the transition matrix elements
to calculate the form factors involved in the radiative
$B\longrightarrow K^{\star}\gamma$ decay. In principle,
we can calculate the form factors at any values of the
squared momentum transfer. In practice, we have made two
approximations, i.e., neglecting the dependence of the kernel on
the relative time and neglecting the contribution of the higher
order negative energy projectors. We obtain a rather genernal
form for the form factors (\ref{a52}) and (\ref{a53}).
These expressions are rather model independent(
apart from the two approximations mentioned above).
We then use a QCD-motivated one gluon exchange plus linear
confinement potential as the kernel to solve the BS equation.
With a rather popular choice for the potential parameters and quark
masses, we get the scalar wave functions $\varphi$ and $f$, and then
calculate the form factors $A$ and $B$. For $\left |V_{ts} \right |
=0.042, \left |V_{tb} \right |=1, m_t=150 GeV$
we find that the
branching ratio
$BR(B\longrightarrow K^{\star}\gamma)= (3.8-4.6)\times 10^{-5}$,
and $R=(10-12) \% $. $R$ may take a slightly larger value
of $(12-14)\%$
if, instead of choosing $m_b=5.12 \ GeV$, a smaller $b$ quark mass
say $m_b=(4.7-4.9) \ GeV$ is used .

Because $B\longrightarrow K^{\star}\gamma$ is a large recoil
process, the relativistic effects are important. There are two
sources of relativistic effects. One is from relativistic
kinematics, which may be seen in (\ref{a52}) and (\ref{a53})
where the form factors are expressed in terms of Dirac spinors,
and where $E_u\gg m_u, E_s\gg m_s$. In (\ref{a52}) and
(\ref{a53}), due to the large recoil momentum of the $K^{\star}$
meson ${ |\stackrel{\rightharpoonup }{k} |}=
\frac {({M_B}^2-{M_{K^{\star}}}^2)}{2M_B} \simeq \frac {M_B}2$, the
overlap integral of the wave functions of $B$ and $K^{\star}$
becomes much smaller than that in the zero recoil limit.
In connection with this , there is another source of relativistic
effects, i.e., the dynamical effect on the meson wave functions.
With the large recoil momentum the overlap is small and therefore
is particularly sensitive to the meson wave functions, which are
determined by inter-quark forces. To see this we have used three
forms for $\varphi$ and $f$ in (\ref{a52}) and (\ref{a53}):
(a) the solutions of BS equation; (b) the solutions of zeroth order
Schr\"odinger equation with the same inter-quark potential;
(c) the Gaussian wave function. We find that the three results
are quite different and the relativistic effects on the wave
functions $\varphi$ and $f$ are indeed important. With (b) and
(c) the obtained decay rates are similar but smaller by more
than a factor of 2 than with (a). This is because, with BS equation
the Breit-Fermi interactions induced by relativistic motion will
broaden the wave functions in momentum space. The important
effect is that the $B$ meson wave function is broadened by
the color magnetic i.e. the hyperfine spin-spin force
and other spin-independent terms induced by
one gluon exchange, which also gives the $B^{\star}-B$ mass
splitting with a right size${}^{\cite{9}}$. With the broadened
wave functions the overlap integral is increased and a larger
decay rate is achieved.
In this connection, the hadronic matrix element involved
in the decay
$B \longrightarrow K^{\star}\gamma$ may indeed provide a test of
some important ingredients in the inter-quark dynamics.
The effects of Breit-Fermi Hamiltonian on the wave functions
have also been shown in charmonium decays${}^{\cite{14}}$.
The observed suppressions for the electric dipole transitions
$\psi'\longrightarrow \gamma\chi_J$ and $\chi_J\longrightarrow
\gamma J/\psi$(J=0,1,2), as well as ${}^1P_1\longrightarrow
\gamma\eta_c$ are probably due to these effects,
because the broadened wave functions in momentum space
(hence narrowed in coordinate space)
by the Breit-Fermi interactions will reduce the dipole transition
overlap integrals, and therefore the rates.

Of course, there are theorectical uncertainties in our approach,
such as the
neglect of the retardation effects
and the contribution of the negative energy projectors,
the lack of knowledge for the
correction
to the static inter-quark potential due to the
light quark motion,
and the gluon(hard and soft)
exchange
nonspectator
effects that are difficult to compute in the quark models at present.
Hopefully, our results can serve as an useful estimate of this
decay.
Definite conclusions strongly depend upon the reduction of these
uncertainties
in the theoretical computations which, we hope, will be improved in
the future.

\bigskip
\bigskip
\centerline{\large\bf ACKNOWLEDGMENTS}

\bigskip
One of the authors(J.T.) would like to thank Prof. D.H. Qin  for
providing him good
conditions for computation. K.T.C. would like to thank Prof. Y.B. Dai
for helpful discussions.
This work was supported in part by the National Natural Science
Foundation of China and the State Education Commission of China.

$Note \ added.$~~After this work was completed and submitted for
publication the $CLEO$ Collaboration reported a result of
the inclusive  $b\rightarrow s\gamma$ decay rate (Ref.[19]).
We have therefore included a brief dissussion for it in our
calculations.

\centerline{\large\bf Figure Captions}

\bigskip
\parindent=0pt
{\bf Figure 1}: Diagram for the meson transition
$\mid P,M\rangle \longrightarrow \mid P^{\prime},M^{\prime}\rangle
$.
Here the quark changes its flavor and momentum via the vertex
$\Gamma$,
while the antiquark remains a spectator.

\bigskip
{\bf Figure 2}: The $B$ meson wave function
$\varphi(\stackrel{\rightharpoonup }{p})$,
defined in Eq.~(\ref{a31})
and normalized by Eq.~(\ref{a30}).
The solid line represents the BS equation
solution by solving Eqs.~(\ref{a26}) and (\ref{a27})
while the dashed line
represents the Schr\"odinger equation solution by solving
Eqs.~(\ref{a26})
and (\ref{a27}) but only keeping the lowest order terms
in the nonrelativistic expression.

\bigskip
{\bf Figure 3}: The $K^{\star}$ meson wave function
$f(\stackrel{\rightharpoonup }{p})$,
defined in Eq.~(\ref{a31})
and normalized by Eq.~(\ref{a30}).
The solid line represents the BS equation
solution by solving Eqs.~(\ref{a28}) and (\ref{a29})
while the dashed line
represents the Schr\"odinger equation solution by solving
Eqs.~(\ref{a28})
and (\ref{a29}) but only keeping the lowest order terms
in the nonrelativistic expression.

\end{document}